\documentclass[pss,fleqn,embeddedheads]{w-art}
\usepackage{times}
\usepackage{w-thm}
\usepackage[]{graphicx,subfigure}
\begin{document}
\DOIsuffix{theDOIsuffix}
\Volume{XX}
\Issue{1}
\Copyrightissue{01}
\Month{01}
\Year{2004}
\pagespan{1}{}
\Receiveddate{\sf zzz} \Reviseddate{\sf zzz} \Accepteddate{\sf
zzz} \Dateposted{\sf zzz}



\title{Evidence for Excitonic Polarons in InAs/GaAs Quantum Dots}


\author{V.\ Preisler\footnote{Corresponding
     author: e-mail: {\sf vanessa.preisler@lpa.ens.fr}}\inst{1}}
\address[\inst{1}]{Laboratoire Pierre Aigrain, Ecole Normale
Sup\'erieure, 24 rue Lhomond, 75231 Paris Cedex 05, France}

\author{T.\ Grange\inst{1}}
\author{R.\ Ferreira\inst{1}}
\author{L.A.\ de\ Vaulchier\inst{1}}
\author{Y.\ Guldner\inst{1}}

\author{F.J.\ Teran\inst{2}}

\address[\inst{2}]{Grenoble High Magnetic Field Laboratory, CNRS/MPI, 25
avenue des Martyrs, 38042 Grenoble Cedex  9, France}

\author{M.\ Potemski\inst{2}}
\author{A.\ Lema\^\i tre\inst{3}}

\address[\inst{3}]{Laboratoire de Photonique et Nanostructures, Route de
Nozay, 91460 Marcoussis, France}

\begin{abstract}
We investigate the interband transitions in several ensembles of
self-assembled InAs/GaAs quantum dots by using photoluminescence
excitation spectroscopy under strong magnetic field.  Well defined resonances are observed in the spectra.
A strong anticrossing between two transitions is observed in all
samples, which cannot be accounted for by a purely excitonic
model.  The coupling between the mixed exciton-LO phonon states is
calculated using the Fr\"ohlich Hamiltonian. The excitonic polaron
energies as well as the oscillator strengths of the interband
transitions are determined.  An anticrossing is predicted when two
exciton-LO phonon states have close enough energies with phonon
occupations which differ by one.  A good agreement is found
between the calculations and the experimental data evidencing the
existence of excitonic polarons.
\end{abstract}
\maketitle                   




\renewcommand{\leftmark}

\section{Introduction}
Various experimental and theoretical works  have demonstrated that
carriers confined in semiconductor quantum dots (QDs) are strongly
coupled to the longitudinal optical (LO) vibrations of the
underlying semiconductor lattice
\cite{hameau02,sarkar05,knipp97,hameau99,verzelen00,inoshita97,li98}.
Far-infrared (FIR) magneto-absorption in $n$- and $p$-doped QDs
have shown that intraband optical transitions involve polaron
levels instead of the purely electronic states
\cite{hameau02,hameau99,preisler05}. Recent theoretical works have
shown that excitons in QDs strongly couple to LO phonons in spite
of their electrical neutrality \cite{verzelen02}. The
eigenstates of the interacting exciton and phonon systems are the
so-called excitonic polarons which are predicted to give
significant modifications of the energy levels and large
anticrossings when two exciton-phonon states have close enough
energies with phonon occupations which differ by one. In the
present work, we have studied, at $T$= 4 K, the
photoluminescence excitation (PLE) under strong magnetic field up
to 28 T of several ensembles of self-assembled InAs/GaAs QDs.
PLE spectroscopy probes the absorption of a subensemble of similar
QDs defined by the detection energy $E_{det}$ and thus allows to circumvent
part of the inhomogeneous broadening caused by dot size inhomogenity. Several well defined
resonances are observed in all samples.  The magnetic field
dependence of the resonance energies allows an unambiguous
assignment of the interband transitions.  A strong anticrossing
between two transitions is observed in all samples as the magnetic
field is changed. Such an anticrossing cannot be accounted for by
a purely excitonic model and one has to consider the
exciton-lattice interactions. We have calculated the coupling
between the mixed exciton-LO phonon states using the Fr\"ohlich
Hamiltonian and we have determined the excitonic polaron states as
well as the energies and oscillator strengths of the interband
transitions.

\begin{figure*}[t]
\begin{center}
\mbox{ \subfigure{
\includegraphics[width=0.30\textwidth]{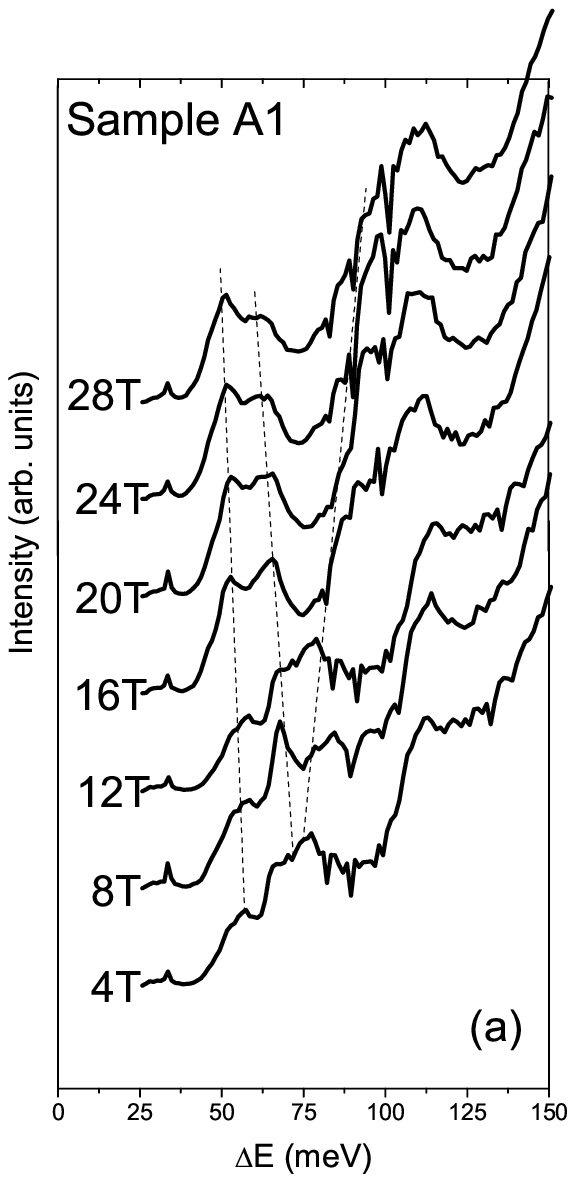}
  \label{fig3a}}

\subfigure{
\includegraphics[width=0.30\textwidth]{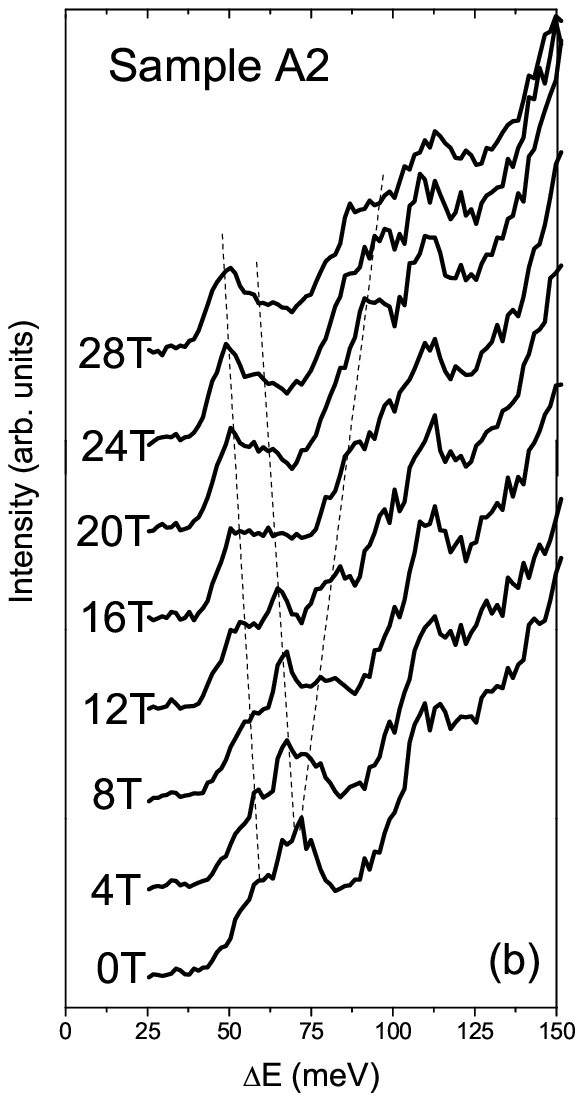}
 \label{fig3b} }
 \subfigure{
\includegraphics[width=0.30\textwidth]{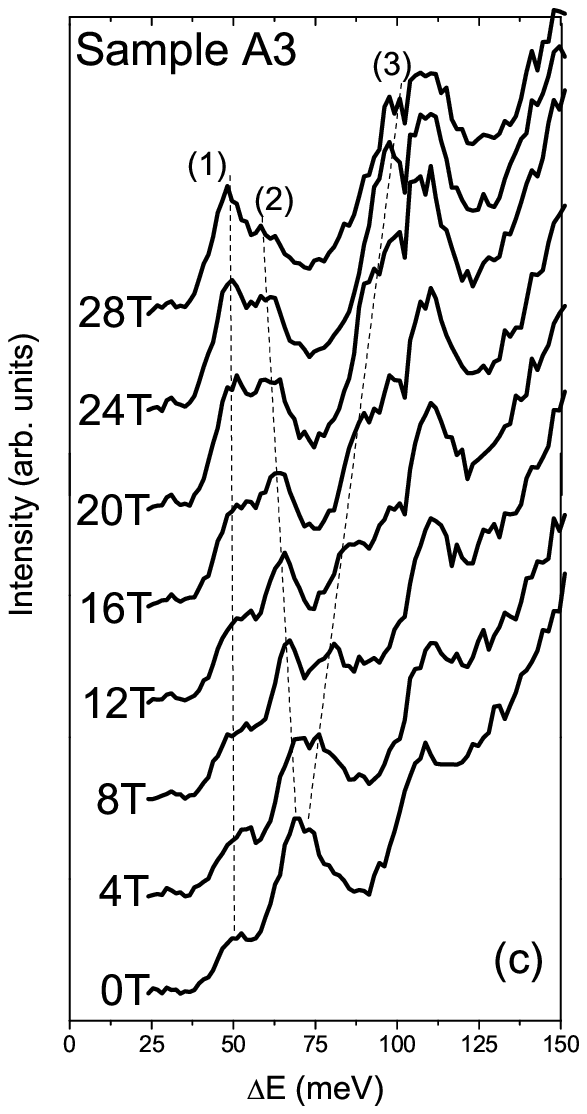}
  \label{fig3c}}}
 \caption{Magneto-PLE spectra of samples A1 (a), A2 (b) and A3 (c) at 4 K from $B$=0
 to $B$=28 T every 4 T and for a $E_{det}$=1215 meV. The dashed lines are guides for the eyes. Traces have been vertically offset for clarity.}
 \label{ple}
\end{center}
\end{figure*}

\section{PLE Results}

\begin{figure}[t]
\begin{center}
\includegraphics[width=0.50\textwidth]{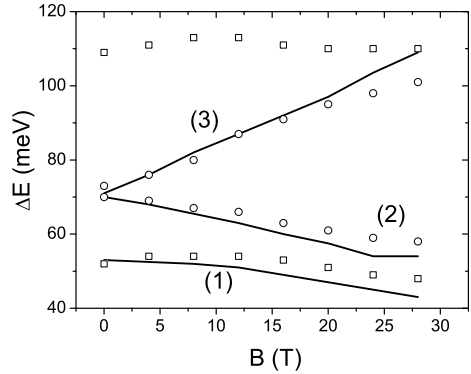}
\caption{Magnetic field dispersion of PLE resonances in sample A3
in open figures with the calculated energy transitions in solid
lines.} \label{fig4}
\end{center}
\end{figure}

Figure~\ref{ple} depicts the PLE
spectra of the three samples A1, A2 and A3 which were respectively undoped, $n$-doped and $p$-doped. The peaks observed  are
associated with transitions between bound levels in the QDs: a low-energy peak is observed at $\sim$50 meV,
a second peak is found at $\sim$75 meV, a third peak is found at
$\sim$110 meV.  For
QDs with a perfect cylindrical symmetry, the ground and first
excited states in both the conduction and valence band are $s$ and
$p$-like respectively. In order to associate the excitation peaks
with transitions in the QDs, a magnetic field $B$ is applied along
the sample's growth axis, as it is well known that the effect of a
magnetic field is different for $s$, $p$ and $d$ states. The energy of the PLE peaks as a function of magnetic field
is plotted in Fig.~\ref{fig4} for sample A3.  As the magnetic
field increases, the peak that was initially at $\sim$ 75 meV splits into
two separate peaks: one peak that increases in energy and a second
peak that decreases in energy. These peaks can be associated with
$p$-like transitions.
Let us now take a closer look at the oscillator strengths of the
lower energy peaks, as shown in Fig.~\ref{ple}.  At 0 T, the
intensity of the low-energy peak at 50 meV is about 20\% of that
of the 75 meV peak.  As the magnetic field increases, an exchange
of oscillator strength between these two peaks is observed. For
samples A1 and A3, at 20 T the two peaks have the same intensity
and by 24 T the oscillator strength of the low-energy peak has
surpassed that of its neighbor. For sample A2, the energy
difference between the two peaks at 0T is smaller and therefore
the anticrossing is observed at a lower magnetic field.  Such a
behavior can not be explained using a purely electronic model. It
is necessary to use a model that takes into account the coupling
between the optical phonons and the photo-created electron-hole
pair in the QD.

\section{Calculation of excitonic polaron}

The noninteracting exciton-phonon states
are labelled $|n_{e},n_{h},N_{\textbf{q}}\rangle$, where
$|n\rangle=|s\rangle,|p^{\pm}\rangle$ are purely electronic
levels. $|N_{\textbf{{q}}}\rangle$ denotes the ensemble of the $N$
LO-phonon states in the $\{\textbf{q\}}$ modes. The Fr\"{o}hlich
Hamiltonian $V_F$ couples states which differ by one phonon.

\begin{equation}
 \langle n_e,n_h,0|V_{F}|n'_e,n'_h,1_{\textbf{q}}\rangle = \frac{A_F}{q}
  \big[ \delta_{n_hn'_h}v_{n_en'_e}(\textbf{q})    - \delta_{n_en'_e} v_{n_hn'_h}(\textbf{q}) \big]
\end{equation}
where $ v_{nn'}(\textbf{q}) = \langle
n|e^{i\textbf{q}.\textbf{r}}|n'\rangle $. The term $A_F$ includes
in its definition the Fr\"{o}hlich constant which is taken as
$\alpha = 0.11$, consistent with precedent polaron studies in
QDs \cite{hameau02}.

\begin{figure}
\begin{center}
\mbox{
\subfigure{\includegraphics[width=0.60\textwidth]{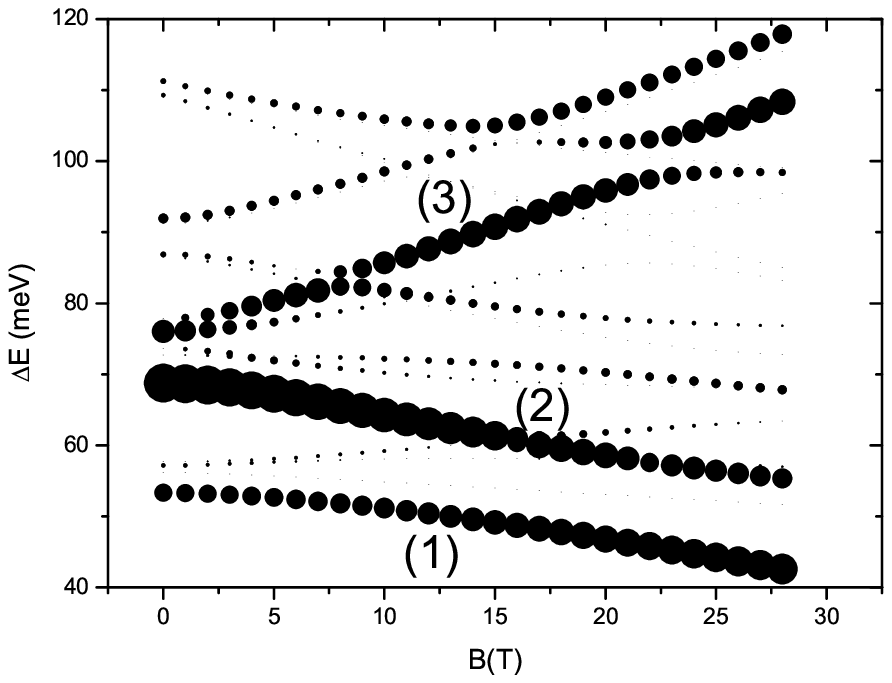}
}
\subfigure{\includegraphics[width=0.35\textwidth]{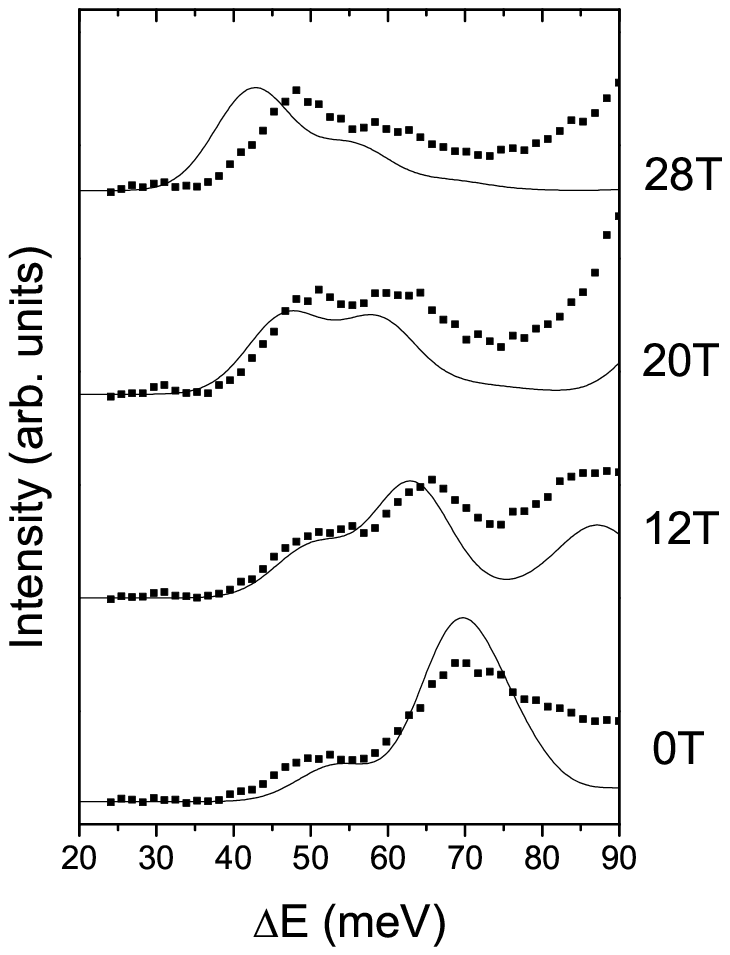}}
}
\end{center}
\hbox{\hspace{0.50in}
\parbox{2.6in}{\begin{small}\textbf{Fig. 3} Calculated excitonic polaron energies and intensities as
a function of the magnetic field. The area of the circles are proportional to the oscillator strengths.\end{small}}
\hspace{0.5in}
\parbox{2.25in}{\begin{small}\textbf{Fig. 4} Experimental (full squares) and calculated (solid lines)
spectra for different magnetic fields.  The experimental data was
taken for $E_{det}$=1215 meV and for sample A3.\end{small}} }

\end{figure}

In order to calculate the excitonic polarons, we have used the
dispersionless LO phonon approximation, which allows the very
accurate calculation of polaron levels in QDs \cite{hameau02}. An
LO phonon energy of $\hbar\omega$ = 36 meV has been used. In
addition to Fr\"{o}hlich coupling terms, coulomb interactions and
in-plane anisotropy have been taken into account. The energy
dispersions and oscillator strengths of the polaron states,
presented in Fig.~3, were calculated using an in-plane effective
mass for electrons $m_{e}$= 0.07$m_{o}$ and for holes $m_{h}$=
0.22$m_{o}$, in agreement with FIR intraband magnetospectrocopy
results.  The other dots parameters were chosen in order to fit
the experimental values of both the intraband $s$-$p$ electronic
(hole) transitions [47 meV (22 meV)] and the interband ground
state energy (detection energy of $1215$ meV). More details on
polaron calculation can be found in reference \cite{preisler06}.
The numerical diagonalization of the full Hamiltonian gives us the
polaron energy transitions. The energy positions as a function of
magnetic field are in good agreement with our data, as seen in
Fig.~\ref{fig4}~\cite{highenergy}. In particular, the Fr\"ohlich
coupling between the states $|p^{-}_e,p^{+}_h,0\rangle$ and
$|s_{e},p^{+}_h,1\rangle$ is responsible for the experimentally
observed strong anticrossing. We calculate also the interband
absorptions for an inhomogeneous ensemble of dots. The solid lines
in Fig.~4 represent the calculated absorption spectra at different
magnetic fields. We compare our calculated interband absorption
spectra with the PLE experimental data. The full squares are data
points taken for sample A3. The evolution of the oscillator
strengths of the absorptions with the magnetic field is very well
described by our model. We are able to predict the exchange of
oscillator strength observed in our results demonstrating the
validity of our analysis and the existence of excitonic polarons.
Similar agreement is found for results obtained for sample A1 and
A2.

\section{Conclusion}
We have investigated the interband transitions in
several ensembles of self-assembled InAs/GaAs QDs by using PLE
spectroscopy.  The magnetic field
dependence of the interband transitions allows their unambiguous
assignment.  When two exciton-LO phonon
states have close enough energies with phonon occupations which
differ by one, a large anticrossing is theoretically predicted.
Such a situation is experimentally induced in our samples by the
applied magnetic field for the two interband transitions
($|p^{-}_e,p^{+}_h,0\rangle,|s_{e},p^{+}_h,1\rangle$)
and a strong anticrossing is actually observed in all the
investigated samples.  We have calculated the coupling between the
mixed exciton-LO phonon states using the Fr\"ohlich Hamiltonian
and we have determined the energies and oscillator strengths of
the interband transitions. Our model accounts well for the
experimental data, evidencing that the excitons and LO-phonons are
in a strong coupling regime in QDs and the interband transitions
occur between excitonic polaron states.  Finally, we believe that
the existence of excitonic polarons could present important
consequences for the energy relaxation in excited QDs and for the
coherence decay times of the fundamental optical transitions.

\begin{acknowledgement}
The Laboratoire Pierre Aigrain is a ``Unit\'e Mixte de Recherche"
(UMR 8551) between Ecole Normale Superieure, the University Pierre
et Marie Curie (Paris 6) and the CNRS. We would like to thank G.
Bastard and S. Hameau for very valuable and fruitful discussions.
\end{acknowledgement}


\begin{thebibliography}{10}
\bibitem{highenergy}
The peak at $\sim$ 110 meV is not predicted in our calculation. A
possible explanation for this peak could be the presence of
confined light holes states, which we do not inlcude in our
calculation.
\end{thebibliography}
\end{document}